\documentclass[prd,reprint,nofootinbib,notitlepage,aps,tightenlines,
preprintnumbers,amsmath,amssymb,showpacs,superscriptaddress]{revtex4-1}

\pdfoutput=1

\usepackage[hyperindex,breaklinks]{hyperref}
\usepackage{graphicx,epstopdf,amsmath,amsfonts,amssymb,color}

\def\be{\begin{equation}}
\def\ee{\end{equation}}
\def\ba{\begin{eqnarray}}
\def\ea{\end{eqnarray}}
\def\ge{\mathrel{\raise.3ex\hbox{$>$\kern-.75em\lower1ex\hbox{$\sim$}}}}
\def\la{\mathrel{\raise.3ex\hbox{$<$\kern-.75em\lower1ex\hbox{$\sim$}}}}

\def\simgt{\mathrel{\raise.3ex\hbox{$>$\kern-.75em\lower1ex\hbox{$\sim$}}}}
\def\simlt{\mathrel{\raise.3ex\hbox{$<$\kern-.75em\lower1ex\hbox{$\sim$}}}}

\newcommand{\fr}[2]{\frac{#1}{#2}}

\newcommand{\nc}{\newcommand}

\nc{\gone}{\bar g_{\pi NN}^{(1)}}
\nc{\gzero}{\bar g_{\pi NN}^{(0)}}
\nc{\al}{\alpha}
\nc{\ga}{\gamma}
\nc{\de}{\delta}
\nc{\ep}{\epsilon}
\nc{\ze}{\zeta}
\nc{\et}{\eta}
\nc{\ka}{\kappa}
\nc{\rh}{\rho}
\nc{\si}{\sigma}
\nc{\ta}{\tau}
\nc{\up}{\upsilon}
\nc{\ph}{\phi}
\nc{\ch}{\chi}
\nc{\ps}{\psi}
\nc{\om}{\omega}
\nc{\Ga}{\Gamma}
\nc{\De}{\Delta}
\nc{\La}{\Lambda}
\nc{\Si}{\Sigma}
\nc{\Up}{\Upsilon}
\nc{\Ph}{\Phi}
\nc{\Ps}{\Psi}
\nc{\Om}{\Omega}
\nc{\ptl}{\partial}
\nc{\del}{\nabla}
\nc{\ov}{\overline}
\nc{\newcaption}[1]{\centerline{\parbox{15cm}{\caption{#1}}}}
\nc{\us}{U(1)$_S$}
\nc{\Rg}{$R_{\gamma\gamma}$}

\def\beq{\begin{equation}}
\def\eeq{\end{equation}}
\def\bmat{\begin{displaymath}}
\def\emat{\end{displaymath}}
\def\bear{\begin{eqnarray}}
\def\eear{\end{eqnarray}}
\def\ba{\begin{eqnarray}}
\def\ea{\end{eqnarray}}
\def\bery{\begin{array}}
\def\ery{\end{array}}
\def\bit{\begin{itemize}}
\def\eit{\end{itemize}}
\def\ben{\begin{enumerate}}
\def\een{\end{enumerate}}
\def\btab{\begin{tabular}}
\def\etab{\end{tabular}}
\def\btbl{\begin{table}}
\def\etbl{\end{table}}
\def\bfig{\begin{figure}[tb]}
\def\efig{\end{figure}}
\def\bpic{\begin{picture}}
\def\epic{\end{picture}}

\def\ga{\mathrel{\raise.3ex\hbox{$>$\kern-.75em\lower1ex\hbox{$\sim$}}}}
\def\la{\mathrel{\raise.3ex\hbox{$<$\kern-.75em\lower1ex\hbox{$\sim$}}}}
\def\gappeq{\mathrel{\rlap {\raise.5ex\hbox{$>$}}
{\lower.5ex\hbox{$\sim$}}}}
\def\lappeq{\mathrel{\rlap{\raise.5ex\hbox{$<$}}
{\lower.5ex\hbox{$\sim$}}}}

\def\gyr{{\rm \, G\kern-0.125em yr}}
\def\mev{{\rm \, Me\kern-0.125em V}}
\def\gev{{\rm \, Ge\kern-0.125em V}}
\def\tev{{\rm \, Te\kern-0.125em V}}

\def\lsim{\mathrel{\rlap{\lower4pt\hbox{\hskip1pt$\sim$}}
    \raise1pt\hbox{$<$}}}                
\def\gsim{\mathrel{\rlap{\lower4pt\hbox{\hskip1pt$\sim$}}
    \raise1pt\hbox{$>$}}}                

\newcommand{\hef}{\ensuremath{{}^4\mathrm{He}}}
\newcommand{\het}{\ensuremath{{}^3\mathrm{He}}}
\newcommand{\lisx}{\ensuremath{{}^6\mathrm{Li}}}
\newcommand{\lisv}{\ensuremath{{}^7\mathrm{Li}}}
\newcommand{\bes}{\ensuremath{{}^7\mathrm{Be}}}
\newcommand{\deut}{\ensuremath{\mathrm{D}}}
\newcommand{\trit}{\ensuremath{\mathrm{T}}}

\newcommand{\hyd}{\ensuremath{\mathrm{H}}}

\newcommand{\keV}{\ensuremath{\mathrm{keV}}}
\newcommand{\MeV}{\ensuremath{\mathrm{MeV}}}
\newcommand{\GeV}{\ensuremath{\mathrm{GeV}}}

\renewcommand{\sec}{\ensuremath{\mathrm{s}}}

\begin{document}

\title{Cosmological Constraints on Very Dark Photons}

\author{Anthony Fradette}
\affiliation{Department of Physics and Astronomy, University of Victoria, 
Victoria, BC V8P 5C2, Canada}

\author{Maxim Pospelov}
\affiliation{Department of Physics and Astronomy, University of Victoria, 
Victoria, BC V8P 5C2, Canada}
\affiliation{Perimeter Institute for Theoretical Physics, Waterloo, ON N2J 2W9, 
Canada}

\author{Josef Pradler}
\affiliation{Institute of High Energy Physics, Austrian Academy of
  Sciences, A-1050 Vienna, Austria}

\author{Adam Ritz}
\affiliation{Department of Physics and Astronomy, University of Victoria, 
Victoria, BC V8P 5C2, Canada}

\date{July 2014}

\begin{abstract}
  We explore the cosmological consequences of kinetically mixed dark photons with a
  mass between 1~MeV and 10~GeV, and an effective electromagnetic
  fine structure constant as small as $10^{-38}$. We calculate the freeze-in
  abundance of these dark photons in the early Universe and explore the
  impact of late decays on BBN and the CMB. This leads to new
  constraints on the parameter space of mass $m_V$ vs kinetic mixing parameter $\kappa$.
\end{abstract}
\maketitle

\section{Introduction}

In the past two decades, there has been impressive progress in our
understanding of the cosmological history of the Universe.  A variety of
precision measurements and observations point to a specific sequence of
major cosmological events: inflation, baryogenesis, big bang
nucleosynthesis (BBN), recombination and the decoupling of the cosmic
microwave background (CMB). While our knowledge of inflation and baryogenesis, likely
linked to the earliest moments in the Universe, is necessarily more uncertain, BBN and the CMB have 
a firm position in cosmic chronology. This by itself puts many models of particle physics to a
stringent test, as the increasing precision of cosmological data leaves
less and less room for deviations from the minimal scenario of standard
cosmology.  In this paper, we adhere to the standard cosmological model, taking as given the above 
sequence of the main cosmological events. Thus we assume that the
Universe emerged from the last stage of inflation and baryogenesis well before the onset of BBN. 
These minimal assumptions will allow us to set stringent bounds on very weakly interacting sectors 
of new physics beyond the Standard Model (SM).

Neutral hidden sectors, weakly coupled to the Standard Model, are an intriguing possibility for new physics. They are motivated 
on various fronts, e.g. in the form of right-handed neutrinos allowing for neutrino oscillations, or by the need for non-baryonic dark matter. 
While the simplest hidden sectors in each case may consist of a single state, various extensions have been explored in recent
years, motivated by specific experimental anomalies. In particular, these extensions allow for models of dark matter 
with enhanced or suppressed interaction rates or sub-weak scale masses. 

From a general perspective, we would expect leading couplings to a
neutral hidden sector to arise through relevant and marginal
interactions. There are only three such flavor-universal `portals' in
the SM: the relevant interaction of the Higgs with a scalar operator
${\cal O}_SH^\dagger H$; the right-handed neutrino coupling $LHN_R$;
and the kinetic mixing of a new U(1) vector $V_\mu$ with hypercharge
$B_{\mu\nu} V^{\mu\nu}$. Of these, the latter vector portal is of
particular interest as it leads to bilinear mixing with the photon and
thus is experimentally testable, and at the same time allows for a
vector which is naturally light. This portal has been actively studied
in recent years, particularly in the `dark force' regime in which the
vector is a loop factor lighter than the weak scale,
$m_V\sim$~MeV--GeV \cite{Essig:2013lka}.

The model for this hidden sector is particularly simple. 
Besides the usual kinetic and mass terms for $V$, the coupling to the SM is
given by \cite{Holdom:1985ag}
\be
 {\cal L}_{\rm V} = -\frac{\kappa}{2} F_{\mu\nu} V^{\mu\nu}  = e \kappa V_\mu J^\mu_{\rm em}.
 \label{LV}
 \ee Thus all phenomenological consequences of the model, including
 the production and decay of new vectors, are regulated by just two
 parameters, $\kappa$ and $m_V$. This makes the model a very simple
 benchmark for all light, weakly interacting, particle searches. There
 are, however, options with regard to the origin of the mass of $V$,
 either a new Higgs mechanism, or $m_V$ as a fundamental
 parameter---the so-called Stueckelberg mass. In this paper, we will
 concentrate on the latter option for simplicity.

 The SM decay channels of $V$ are well known. In the mass range where
 hadronic decays are important, one can use direct experimental data
 for the $R$-ratio to infer couplings to virtual time-like photons,
 and hence to determine the decay rate $\Gamma_V$ and all the
 branching ratios. In a wide mass range from $\sim 1 - 220$ MeV, the
 vector $V$ decays purely to electron-positron pairs with lifetime \be
\label{lifetime}
\tau_V \simeq  \frac{3}{\alpha_{\rm eff} m_V }  = 
6\times 10^5  {\rm \;yr} \times \frac{10\,{\rm MeV} }{ m_V} \times \frac{10^{-35}}{\alpha_{\rm eff} }
\ee
where we have introduced the effective electromagnetic fine structure constant, absorbing the square of the mixing angle into 
its definition,
\be
 \al_{\rm eff} \equiv \al \ka^2. 
 \ee
Importantly, we assume no light hidden sector states $\chi$ charged under U(1), so that there are no 
``dark decays" of $V\to \chi \bar\chi$ that would erode the visible modes and shorten the lifetime of $V$. 

The normalization of the various quantities in (\ref{lifetime}) roughly identifies the region of interest in the $\{\kappa,m_V\} $
parameter space for this paper. We will explore the cosmological consequences of 
these hidden U(1) vectors with masses in the MeV-GeV range,
and lifetimes long enough for the decay products to directly influence the physical processes in the universe following BBN, and 
during the epoch of CMB decoupling. These vectors have a parametrically small
coupling to the electromagnetic current, and thus an extremely small production cross sections for $e^+e^-\to V\gamma$,
\be
 \sigma_{\rm prod} \sim \frac{\pi \alpha \alpha_{\rm eff}}{E_{\rm c.m.}^2} \sim
 10^{-66}-10^{-52} ~ {\rm cm}^2, \label{prod}
\ee
where we took $E_{\rm c.m.}\sim 200$ MeV and the range is determined by our region of interest,
\be
 \al_{\rm eff} \sim 10^{-38} - 10^{-24}.
\ee
Such small couplings render these vector states completely undetectable in terrestrial
particle physics experiments, and consequently we refer to them as {\it very dark photons} (VDP). 
As follows from the expression (\ref{lifetime})  for the lifetime, the lower limit of the above range for $\al_{\rm eff}$ is relevant for CMB physics, while the upper 
limit is important for BBN. 

The production cross section (\ref{prod}) looks prohibitively small, but in the early Universe at $T\sim m_V$ every particle in the primordial plasma 
has the right energy to emit $V$'s. The cumulative effect of early Universe production at these temperatures,
followed by decays at $t \sim \tau_V$, can still inject a detectable amount of electromagnetic energy. 
A simple parametric estimate for the electromagnetic energy release per baryon, omitting ${\cal O}(1)$ factors, takes the form
\begin{equation}
\label{estimate}
E_{\rm p.b.} \sim \frac{m_V\Gamma_{\rm prod} H^{-1}_{T=m_V}} {n_{b,T=m_V}}\sim \frac{\alpha_{\rm eff} M_{\rm Pl}}{10\, \eta_b} 
\sim \alpha_{\rm eff} \times 10^{36}\,{\rm eV}.
\end{equation}
Here the production rate per unit volume, $\Gamma_{\rm prod}$, was taken to be the product of the 
typical number density of particles in the primordial plasma and the $V$ decay rate, $ \tau_V^{-1} n_{\gamma,T=m_V}$.
This production rate is active within one Hubble time, $H^{-1}_{T=m_V}$, leading to the appearance of the Planck mass in (\ref{estimate}), along with another large factor, the ratio of photon to baryon number densities, $\eta_b^{-1} =1.6 \times 10^9$. 
One observes that the combination of these two factors is capable of overcoming the extreme suppression by $\alpha_{\rm eff}$.
Given that BBN can be sensitive to an energy release as low as $O({\rm MeV})$ per baryon, and that the CMB anisotropy spectrum allows 
us to probe sub-eV energy injection, we reach the conclusion that the early Universe can be an effective probe of VDP!
The cosmological signatures of the decaying VDP were partially explored in \cite{Redondo:2008ec,Pospelov:2010cw},
but to our knowledge the CMB constraints on this model were not previously studied. 

\begin{figure}[htb]
\begin{center}
\includegraphics[width=\columnwidth]{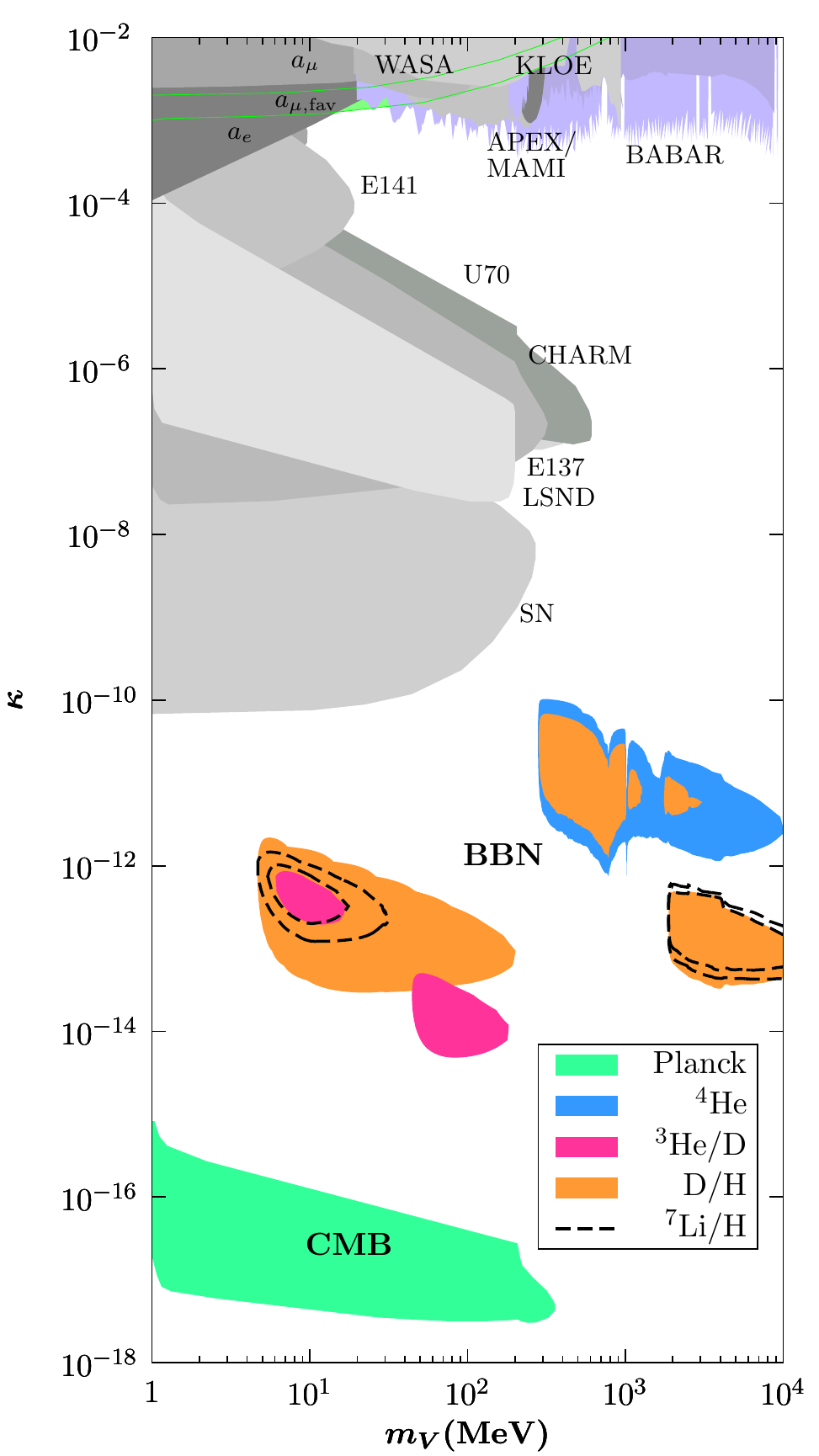}
\caption{An overview of the constraints on the plane of vector mass versus kinetic mixing, showing the regions excluded
due to their impact on BBN and the CMB anisotropies, in addition to various terrestrial limits \cite{Essig:2013lka,snowmassrefs}, including the more recent limits \cite{recentlimits}. These excluded regions are shown in more detail in later sections.}
\label{fig:sum}
\end{center}
\end{figure}

In the remainder of this paper, we provide detailed calculations to
delineate the VDP parameter regions that are constrained by BBN and
CMB data.  In the process, we provide in Section~2 an improved
calculation of the `freeze-in' abundance in the Early Universe (using
some recent insight about the in-medium production of dark vectors
\cite{An:2013yfc,An:2013yua}; see also~\cite{Redondo:2013lna}).  In
Section~3, we explore the BBN constraints in more detail, including
the speculative possibility that the currently observed over-abundance
of $^7$Li can be reduced via VDP decays. Then in Section~4 we consider
the impact of even later decays on the CMB anisotropies. A summary of
the constraints we obtain in shown in Fig.~\ref{fig:sum}, and more
detailed plots of the parameter space are shown in Sections~3 and
4. We finish with some concluding remarks in Section~5. Several
Appendices contain additional calculational details.

\section{Freeze-in abundance of VDP} 

\begin{figure}[htb]
\begin{center}
\includegraphics[width=.35\textwidth]{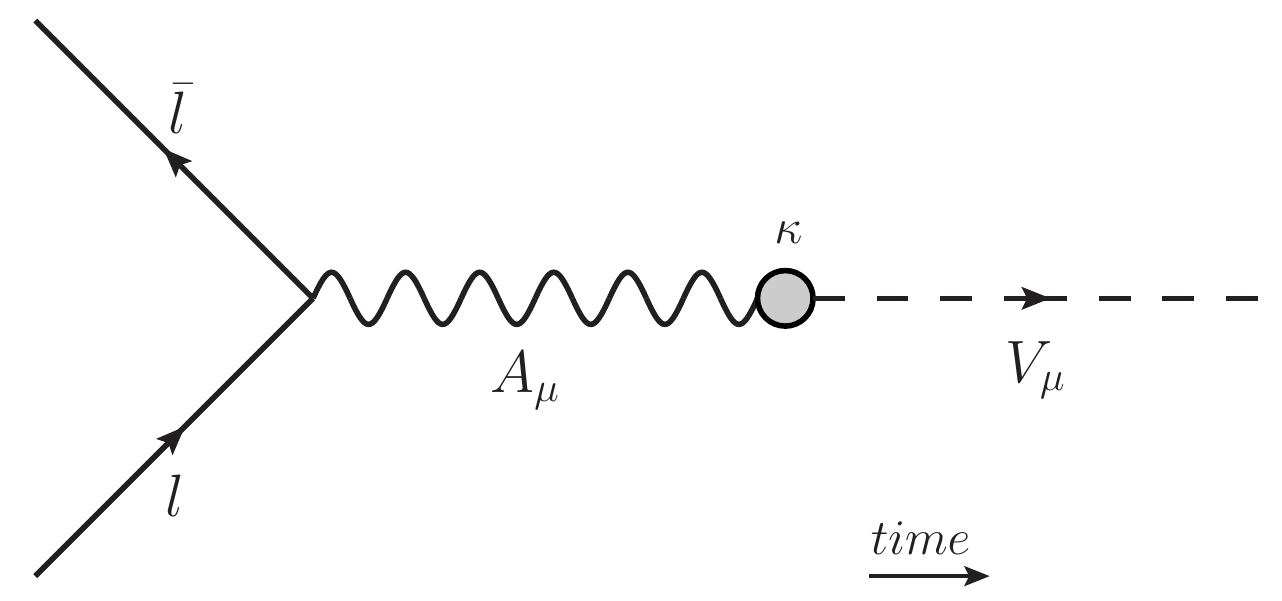}
\caption{Illustration of the coalescence production of the dark photon $V$ via an off-shell photon.}
\label{fig:diagram-eebartoV}
\end{center}
\end{figure}

The cosmological abundance of long-lived very dark photons is determined by the freeze-in mechanism. While there are several 
possible production channels,  the simplest and most dominant is  the inverse decay process. 
When quark (or more generally hadronic) contributions can be neglected, 
the inverse decay proceeds via coalescence of  $e^\pm$ and $\mu^\pm$, $l\bar{l}\to V$, shown in
Fig.~\ref{fig:diagram-eebartoV}.

The Boltzmann equation for the total number density of $V$ takes the form
\begin{eqnarray}
\label{Beq}
\dot{n}_V +3Hn_V &=& 
\prod_{i=l,\bar{l},V} \int \left(\frac{d^3 {\bf p}_i}{(2\pi)^3 2E_i}\right) N_{l}N_{\bar l}\\
 && \!\!\!\!\!\!\!\!\!\!\! 
(2\pi)^4\delta^{(4)}(p_l+p_{\bar{l}}-p_V) \sum
|M_{l\bar{l}}|^2, \nonumber
\end{eqnarray}
where the right hand side assumes the rate is sub-Hubble so that $V$ never achieves an equilibrium density.  
The product of Fermi-Dirac (FD) occupation numbers, $N_{l(\bar l)} = [1+\exp(-E_{l(\bar l)}/T)]^{-1}$, is usually considered 
in the Maxwell-Boltzmann (MB) limit, $N_{l}N_{\bar l}\to e^{(E_l+E_{\bar l})/T}$.
Although this is not justified parametrically, numerically the FD$\to$MB substitution is quite accurate, because 
as it turns out the peak in the production rate (relative to entropy) is at $T<m_V$ \cite{Redondo:2008ec}.

The matrix element $\sum|M_{l\bar{l}}|^2$ is summed over both initial and final state spin degrees of freedom. 
In general, it should include the in-medium photon propagator in the thermal bath, and the fermion wave functions. 
Among these modifications the most important ones are those that lead to the resonant production of dark photon states.
However, resonant production occurs at much earlier times \cite{Redondo:2008ec}, 
at temperatures $T_{r}^2 \geq 3 m_V^2/ (2 \pi \alpha) \simeq (8 m_V)^2$, and turns out to be 
parametrically suppressed relative to continuum production; the details of the corresponding calculation 
are included in Appendix A. The dominant continuum production corresponds to temperatures of $m_V$ 
and below where the $T$-dependence of $\sum|M_{l\bar{l}}|^2$ can be safely neglected. In the present model it is given by 
\be
\sum |M_{l\bar{l}}|^2 = 16 \pi \alpha_{\rm eff} m_V^2 \left(1+2\frac{m_l^2}{m_V^2}\right). 
\label{eq:M2}
\ee
The same matrix element determines the  decay width,
\be
\Gamma_{V\to l\bar{l}} = \frac{\alpha_{\rm eff}}{3} m_V
\left(1+2\frac{m_l^2}{m_V^2}\right)\sqrt{1-4\frac{m_l^2}{m_V^2}}.
\ee
The right hand side of (\ref{Beq}), that can be understood as the number of $V$ particles emitted per unit volume per
unit time. In the MB approximation, it can be reduced to 
\be
\frac{1}{(2\pi)^3}\frac{1}{4}\int_{\rm Eq.~\ref{intlimits}} dE_l\, dE_{\bar{l}}\, e^{-\frac{E_l+E_{\bar l} }{T}} \sum
|M_{l\bar{l}}|^2,
\ee
where the integration region is given by
\be
\left |\frac{m_V^2}{2} - m_l^2 - E_lE_{\bar l} \right|
\le  \sqrt{E_l^2 - m_l^2}\sqrt{E_{\bar l}^2 - m_l^2}.
\label{intlimits}
\ee 
In the approximation where only electrons are allowed to coalesce and their mass neglected,  
$m_l \ll m_V <2m_\mu$, (\ref{intlimits}) reduces to $E_lE_{\bar{l}}\geq m_V^2/4$ and the
integration leads to the familiar modified Bessel function, 
\be 
\label{simplified}
s\dot{Y}_V = \dot{n}_V +3Hn_V = \frac{3}{2\pi^2} \Gamma_{V\to l\bar{l}} m_V^2 T K_1(m_V/T),
\ee 
where $Y_V=n_V/s$ is the number density normalized by the total entropy density,
and $\Gamma_{V\to l\bar{l}}=\alpha_{\rm eff} m_V/3$, without $(m_l^2/m_V^2)$-suppressed corrections,
is used for consistency. 
The final freeze-in abundance via a given lepton
pair is given by 
\be
Y_{V,f}^l = \int_{0}^\infty dT \frac{\dot{Y}_V^l}{H(T)T}.
\label{eq:Y-Tintegral}
\ee
 The integrals are evaluated numerically using
\be
H(T) \simeq 1.66 \sqrt{g_*(T)}\frac{T^2}{M_{\rm pl}}; \qquad s(T) = \frac{2\pi^2}{45}g_{*}(T)T^3,
\ee
where $g_*(T)$ is the effective number of relativistic degrees of freedom, evaluated with the most recent lattice and perturbative QCD results (see Appendix A for details).

For the simplest case of the MB distribution, with only relativistic electrons and positrons contributing and away from particle thresholds that change $g_*(T)$, 
the final integral can be evaluated analytically, and we have
\begin{equation}
\label{Ysimple}
Y_{V,f}^e = \frac{9}{4\pi} \frac{m_V^3\Gamma_{V\to e\bar{e}} }{(Hs)_{T=m_V}} 
= 0.72 \frac{m_V^3\Gamma_{V\to e\bar{e}} }{(Hs)_{T=m_V}}.
\end{equation}
This number reduces somewhat if the FD statistics is used, $0.72_{\rm MB} \to 0.54_{\rm FD}$, 
but receives a $\sim 20\%$ upward correction from the transverse resonance (see Appendix B). Our numerical integration 
routine includes both the correct statistics and the addition of resonant production. 

While the treatment of leptonic VDP production might be tedious but straightforward, 
hadronic production in the early universe is not calculable in principle, as one cannot simply 
extrapolate measured rates for the conversion of virtual photons to hadrons above temperatures of the 
QCD and/or chiral phase transitions.  While the generic scaling captured by Eq. \ref{Ysimple} holds, one needs to 
make additional assumptions about the treatment of the primordial hadron gas. 
It seems reasonable to assume that at high temperatures, when all light quarks are deconfined, 
the individual quark contribution  $Y_{V,f}^q$ can be added by imposing a lower cutoff at the confinement scale
$T_c$ in the integral~(\ref{eq:Y-Tintegral}) and multiplying the matrix element~(\ref{eq:M2}) by the square of the quark
electric charge $Q_q^2$. Below $T_c$ we will use a free meson gas as an approximation for the hadronic states, and 
production via inverse charged pion and kaon decays $\{\pi^+\pi^- ,K^+K^-\} \to V$ is included using a scalar QED model (see Appendix C).

\begin{figure}[htb]
\begin{center}
\includegraphics[width=.5\textwidth]{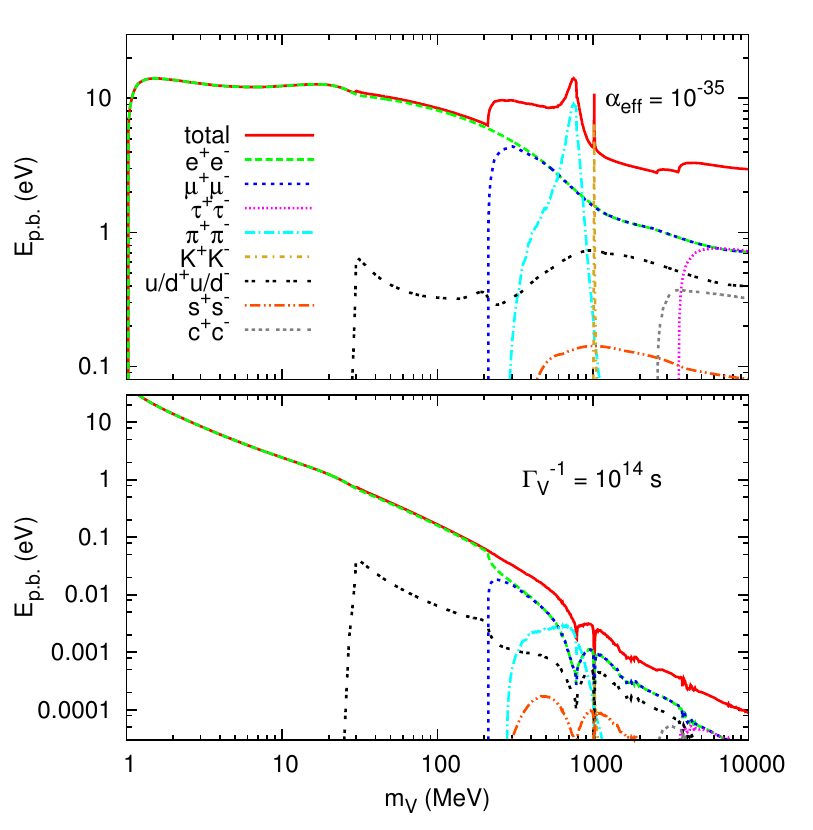}
\caption{Total energy stored per baryons for $\alpha_{\rm eff} = 10^{-35}$ (upper) and $\Gamma_V^{-1} = 10^{14}$s (lower) from the various production channels as labeled.}
\label{fig:Epb}
\end{center}
\end{figure}

The VDPs when produced are semi-relativistic, and the subsequent expansion of the Universe quickly cools them so that 
at the time of decay $E_V = m_V$.  The decay deposits this energy into $e^\pm$, $\mu^\pm$ and $\pi^\pm$ pairs,
and more complicated hadronic final states when $m_V$ is above the $\rho$-resonance. 
Thus, the energy stored per baryon (before the characteristic decay time) is given by 
\be
\label{Eyield} 
E_{\rm p.b.} = m_V Y_{V,f} \frac{s_0}{n_{b,0}},
\ee
where $n_{b,0}/s_0 = 0.9\times 10^{-10}$ is the baryon-to-entropy ratio today. 
$E_{\rm p.b.}$ is shown in two separate panels in Fig.~\ref{fig:Epb}. The top panel 
shows it as a function of $m_V$ at fixed $\alpha_{\rm eff}$, and the lower panel 
 fixes the VDP lifetime to $\tau_V=10^{14}$s. We illustrate the contributions from the different
production channels. Using this calculated VDP energy reservoir 
we are now ready to explore its consequences for BBN and the CMB.

\section{Impact on BBN}
\label{sec:impact-bbn}

Late decays of dark photons affect the epoch of primordial
nucleosynthesis after cosmic time $t\gtrsim 1\,\sec$ in a variety of
ways. The resulting constraints are governed by a combination of
lifetime and abundance, and both have complementary trends with
respect to $m_V$; $\tau_V$ ($Y_V$) decreases (increases) with growing
mass. Therefore we generally expect constraints to be bounded and
localized islands in parameter space with the relevant combination of $
m_V $ and $Y_V$ to ensure BBN sensitivity. 

Prior to decay, $V$ contributes to the matter content substantially,
$n_V/n_b \lesssim 10^{8}$ for $\tau_V < 1\,\sec$. Whereas the modification
of the Hubble rate is generally small, the decays of $V$ imply the
injection of electrons, muons, pions, etc.,~in numbers larger than
 baryons.  The effects on BBN are best described by
partitioning the decay into electromagnetic and hadronic energy
injection and in the following we provide a lightning review of those modes
separately.

MeV-scale dark photons with $m_V < 2 m_{\pi}$ provide a prototypical model
of electromagnetic energy injection because the dominant kinematically
accessible decay modes are $V\to e^+ e^-, \mu^+\mu^-$.  Muons decay
before interacting weakly, and electron-positron pairs are instantly
thermalized via rapid inverse Compton scattering on background
photons.  An electromagnetic cascade forms in energy degrading
interactions of photons and electrons. The large number of photons
created gives rise to a non-equilibrium destruction and creation of
light elements.

The most important feature of the injected photon energy spectrum
$f_{\gamma}(E_{\gamma})$ is a sharp cut-off for energies above the
$e^{\pm}$ pair-creation threshold on ambient photons,
$E_{\mathrm{pair}} \simeq m_e^2/(22T)$. High-energy photons are
efficiently dissipated before they can interact with nuclei, so that
to good approximation $f_{\gamma}(E_{\gamma}) = 0$ for
$E_{\gamma}>E_{\mathrm{pair}}$.  In contrast, less energetic photons
below the pair-creation threshold can interact with the light
elements.  Equating $E_{\mathrm{pair}}$ against the thresholds for
dissociation of the various light elements informs us about the
temperature and hence cosmic time $t_{\rm ph}$ at which to expect the
scenario to be constrained:
\begin{equation*}
  t_{\rm ph} \simeq 
\left\{ \begin{array}{ll@{}r}
2\times 10^4\sec ,&  \ \bes+\gamma\to\het+\hef& (1.59\,\MeV),\\
5\times 10^4\sec  ,&  \ \deut+\gamma\to n+ p & (2.22\,\MeV),\\
4\times 10^6\sec ,& \ \hef+\gamma\to \het/\trit+n/p& (20\,\MeV),
\end{array}\right .
\end{equation*}
where the binding energy of the nucleus against destruction has been
given in brackets.
Finally, note that we also find that neutrino injection from muon
decay does not yield observable changes in the light element
abundances---a facinating story in
itself~\cite{Pospelov:2010cw}.

Once $m_V > 2 m_{\pi}$, the hadronic channels open in the decay of $V$
and the effects on BBN become more difficult to model. A major
simplification is that only long-lived mesons $\pi^{\pm}$, $K^{\pm}$,
and $K_L$, with lifetime $\tau \sim 10^{-8}\,\sec$, and (anti-)nucleons
have a chance to undergo a strong interaction reaction with ambient
protons and nuclei. The relevant reactions are charge exchange,
\textit{e.g.}~$\pi^- + p \to \pi^0 + n $, and absorption with
subsequent destruction of light elements, \textit{e.g.}~$\pi^- + \hef
\to T + n$. Prior to the end of the deuterium bottleneck at $T \simeq
100\,\keV$ only the former reactions are possible. They change the
$n/p$ ratio that determines the primordial $\hef$ value. Later, once
elements have formed, charge exchange creates ``extra neutrons''
on top of the residual and declining neutron abundance. Moreover,
spallation of \hef\ with non-equilibrium production of mass-3
elements and secondaries, \textit{e.g.} through $\trit +\hef_{\rm
  bg}\to \lisx + n$, are important. We model all such reactions in
great detail, include secondary populations of pions from kaon decays,
and various hyperon producing channels from reactions of kaons on
nucleons and nuclei.
A detailed exposition of the hadronic part along with a discussion of
all included reactions can be found in our previous
work~\cite{Pospelov:2010cw}. More details are  provided when 
discussing our findings below as well as in Appendix D.

We now proceed to review the light element observations that form the
basis of our adopted limits.
Probably the most notable recent developments in the determination of
light element abundances are two precision measurements of D/H from
high-$z$ QSO absorption systems~\cite{Pettini:2012ph,Cooke:2013cba}.
Both have error bars that are a factor $\sim 5$ smaller than the
handful of previously available determinations. Taken together, the
mean observationally inferred primordial D/H value now
reads~\cite{Cooke:2013cba},
\begin{align}
\label{eq:cooke}
  {\rm D/H} = (2.53 \pm 0.04) \times 10^{-5} . 
\end{align}
Nonetheless, systematically higher levels of primordial D/H are 
conceivable, in spite of the above error bar. For example, D may
be astrated or absorbed on dust grains. Indeed, values as high as
$4\times 10^{-5}$ have been
reported~\cite{Burles:1997fa,Kirkman:2003uv}, so as a conservative
upper limit we employ,
\begin{align}\label{eq:Dhigh} {\rm D/H} < 3\times 10^{-5} .
\end{align}
On the flip side, underproducing D yields a robust constraint since no
known astrophysical sources of D exist. We account for this constraint
either by adopting the nominal lower $2\sigma$-limit from
(\ref{eq:cooke}) or by demanding,
\begin{align}
\label{eq:hetconstr}
  \het/\deut < 1. 
\end{align}
The latter limit employs the solar system
value~\cite{1993oeep.book.....P} and arises from the consideration
that $\deut$ is more fragile than $\het$, and hence a monotonically
increasing function of time. Despite the uncertain galactic chemical
evolution of $\het$, (\ref{eq:hetconstr}) can therefore be considered
robust.

The inference of the primordial mass fraction $Y_p$ from extragalactic
H-II regions proved to be systematically uncertain in the
past~\cite{Izotov:2010ca,Aver:2013wba} and values in the range
\begin{align}
\label{eq:he4obs}
  0.24\leq Y_p \leq 0.26
\end{align}
have been reported. We adopt this range as our cosmologically viable
region.

Finally, what is believed to be the primordial value of $\lisv/\hyd$,
the so-called Spite plateau~\cite{Spite:1982dd}, is a factor of 3-5
lower than the lithium yield from standard BBN,
$\lisv/\hyd=(5.24^{+0.71}_{-0.67})\times
10^{-10}$~\cite{Cyburt:2008kw}. We deem the lithium problem solved in
this model if we can identify a region in parameter space where lithium
is reduced to the Spite plateau value,
\begin{align}
   10^{-10}<\lisv/\hyd<2.5\times 10^{-5} . 
\end{align}
We take an opportunity to comment that the status of the lithium problem is somewhat 
controversial: while it is possible that new physics is responsible for its solution, the 
astrophysical lithium depletion mechanisms can also be invoked (see ref. \cite{Fields:2011zzb}  for a review of this subject). 

\begin{figure}[htb]
\begin{center}
\includegraphics[width=1.0\columnwidth]{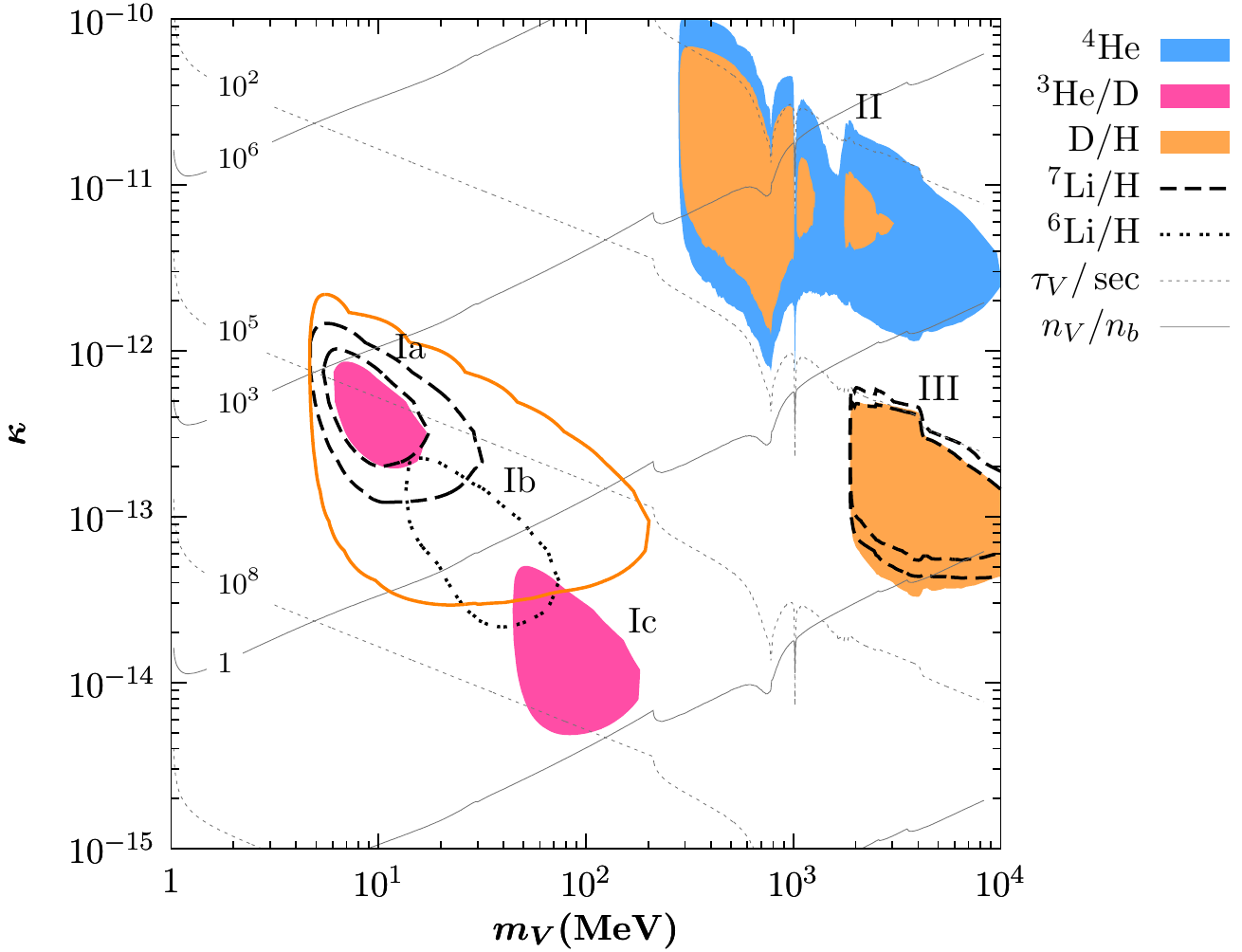}
\caption{Effects on BBN from the decay of relic dark photons as a
  function vector mass of $m_V$ and kinetic mixing parameter
  $\kappa$. The diagonal gray lines are contours of lifetime  $\tau_V$
  (solid) and abundance per baryon $n_V/n_b$ prior to decay
  (dotted). Shaded regions are excluded as they are in conflict with
  primordially inferred light element abundances. The solid (orange)
  closed line is a potential $2\sigma$ constraint from underproduction
  of D/H derived from (\ref{eq:cooke}). The dashed black lines are
  contours of decreasing $\lisv/\hyd$ abundance, $4\times 10^{-10}$
  and $3\times 10^{-10}$, going from the outside to the inside,
  respectively. The dotted line shows $\lisx/\hyd = 10^{-12}$ which
  corresponds to an extra production by about two orders magnitude but
  without being in conflict with observations.}
\label{fig:bbn}
\end{center} 
\end{figure}

We are now in a position to present our results in Fig.~\ref{fig:bbn}
where a scan over the $m_V,\kappa$ parameter space is shown, and
contours of constant lifetime, $\tau_V$ and relic
abundance $n_V/n_b$ prior to decay are shown by the diagonal solid and
dotted lines, respectively.  Three distinct regions labeled I-III are
identified as being in conflict with observations. They arise from
distinct physical processes which we now proceed to describe.

\textbf{Regions I:} 
In the regions labeled I the dark photon exclusively decays to $e^+
e^-$.  They are associated with pure electromagnetic energy injection.

In region Ia with a ballpark lifetime $\tau_V\sim 10^5\,\sec$, \bes\
and D are destroyed. From the outer to the inner (black) dashed
curves, the $\lisv/\hyd$ abundance is reduced to $4\times 10^{-10}$
and $3\times 10^{-10}$ respectively. It is therefore a region in which
the cosmological lithium problem is ameliorated. Smaller abundances of
$\lisv/\hyd$ are disfavored by the constraint $\het/\deut<1$ (pink
shaded region); an equivalent region from the requirement
$\deut/\hyd>10^{-5}$ coincides with this one and is not shown.  If
we take the new measurements~(\ref{eq:cooke}) at face value, the
prospective solution to the lithium problem is excluded altogether
by the nominal $2\sigma$ lower limit on $\deut/\hyd$ shown by the
(orange) solid closed line.

In region Ib, in addition to the potential underproduction of D/H,
photodissociation of $\lisv$ and $\bes$ leads to the primary production
of $\lisx/\hyd> 10^{-12}$. This is not at the level of a constraint,
but we show the dotted contour anyway in order to better
illustrate what is happening in the respective regions of parameter
space. 

Finally, in region Ic, with $V$-lifetime of $\sim 10^7\,\sec$, $\hef$
is being dissociated and the net creation of $\het/\deut$ rules out
this region of parameter space. Once $\hef$ is split, $\lisx$ can be
produced through a secondary mechanism of energetic mass-3 spallation
products such as $\trit + \hef|_{\rm bkg} \to \lisx + n $. We find, however,
that such channels are not efficient enough to provide any additional
constraint.

\textbf{Region II:}
Now we turn to the low-lifetime/high-abundance region~II. The lifetime
of $V$ is below 100~\sec\ and hence marks a choice of parameters where
the dark photon decays before the end of the D-bottleneck ($T\sim
100\,\keV$). The injection of pions and---if kinematically
allowed---of kaons and nucleons, induces $n\leftrightarrow p$
interconversion. It has the general effect that the $n/p$-ratio
rises. The elevated number of neutrons that in turn become
available at the end of the D-bottleneck allow for more D-formation
and subsequently more $\hef$. The region is therefore challenged by
the constraints $Y_p\leq 0.26$ and $\deut/\hyd \leq 3\times 10^{-5}$.

\textbf{Region III:}
Finally, region III is characterized by the presence of ``extra
neutrons'' that appear right after the main stage of nucleosynthesis
reactions at cosmic times $t\sim 10^3\,\sec$. The origin of those
neutrons is twofold. First, there is a direct injection of $n$ from
the decay $V\to n \bar n$. Second, there is indirect production, 
from charge exchange of $\pi^-$ on protons, $\pi^- p\to n\pi^0 $ or
$\pi^- p\to n \gamma$, and from hyperon production by ``s-quark''
exchange of $K^-$ on protons with subsequent hyperon decay. We note in
passing that $K^- p \to \bar K^0 n$ has positive $Q$-value and is not
allowed for stopped kaons; conservatively, we neglect this reaction.

The elevated neutron abundance leads to a chain of reactions that
depletes the overall lithium abundance, 
\begin{align}
 \text{step 1:}&\quad \bes + n \to \lisv + p , \\
   \text{step 2:}&\quad 
 \lisv + p  \to \hef + \hef . 
\end{align}
In the first step, $\bes$ charge exchanges with the neutron and forms
\lisv. In a second step, $\lisv $, because it has one less unit of charge, 
is more susceptible to being destroyed by protons. The result of this
mechanism is shown in Fig.~\ref{fig:bbn} by the dashed curves. Most
of the extra neutrons are, however, intercepted by protons so that
this potential solution to the lithium problem is always accompanied
by an elevated D-yield. The D/H constraint~(\ref{eq:Dhigh}) is shown
by the (orange) solid region.

A more detailed description of the calculations used to obtain these results is provided in Appendix D.

\section{Impact on the CMB}

Later decays of VDP, which occur after recombination if $\tau_V
\gtrsim 10^{13}$s, can leave an imprint on the CMB. In particular, as
discussed in \cite{Chen:2003gz,Zhang:2007zzh}, the altered ionization
history tends to enhance the TE and EE spectra on large scales, while
the TT temperature fluctuation is damped on small
scales. Consequently, precision CMB data can be used to further
constrain the VDP parameter space in regimes where the late decays
impact the ionization history.

\begin{figure}[htb]
\begin{center}
\includegraphics[width=.5\textwidth]{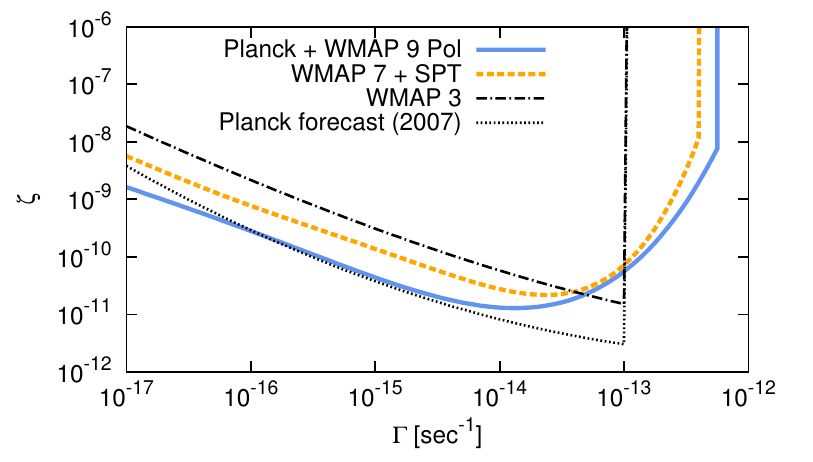}
\caption{CMB constraints on the energy injection parameters $\zeta$ and $\Gamma$.  For comparison, we include the WMAP3 curve and the Planck forecast (2007) from Ref.~\cite{Zhang:2007zzh}.}
\label{fig:cmb-generic-constraints}
\end{center}
\end{figure}

The energy injection of a decaying species can be
generically parametrized as~\cite{Chen:2003gz,Zhang:2007zzh}
\be
 \frac{dE}{dtdV} = 3 \zeta m_{\rm p} \Gamma e^{-\Gamma t},
\label{eq:energ-inject}
\ee 
with $(1-x_e)/3$ of this energy going to ionization and $(1+2x_e)/3$ heating the medium, $x_e$ representing the ionized
fraction. The energy output of each decay is $3\zeta m_{\rm p}$, the normalization chosen so that
(\ref{eq:energ-inject}) gives the ionizing energy after recombination ($x_e\to 0$). Using
\textsc{Class}~\cite{Blas:2011rf} to obtain the CMB power spectra and \textsc{MontePython}~\cite{Audren:2012wb} as
a Monte Carlo Markov Chain driver, we determine the $2\sigma$ limits from the Planck 2013 results~\cite{Ade:2013kta} (which also incorporates the low-$l$ polarization likelihood from WMAP9~\cite{Bennett:2012zja}). The limits are shown in Fig.~\ref{fig:cmb-generic-constraints}, with constraints similarly derived from WMAP7~\cite{Komatsu:2010fb} +
SPT~\cite{Keisler:2011aw}, along with the WMAP3 and 2007 Planck forecast fits from~\cite{Zhang:2007zzh}. The cutoff at $\Gamma^{-1} = 10^{13}$s appears
since Ref.~\cite{Zhang:2007zzh} used a purely matter-dominated approximation for the elapsed time $\left[t(z)\sim
(1+z)^{-3/2}\right]$ in the exponential of~(\ref{eq:energ-inject}) and assumed that decay lifetimes shorter than
$10^{13}$ seconds happen before recombination and do not impact the CMB. In our calculations, we use the exact time
from $\Lambda$CDM cosmology and obtain a more accurate picture for shorter lifetimes.

\begin{figure}[htb]
\begin{center}
\includegraphics[width=.5\textwidth]{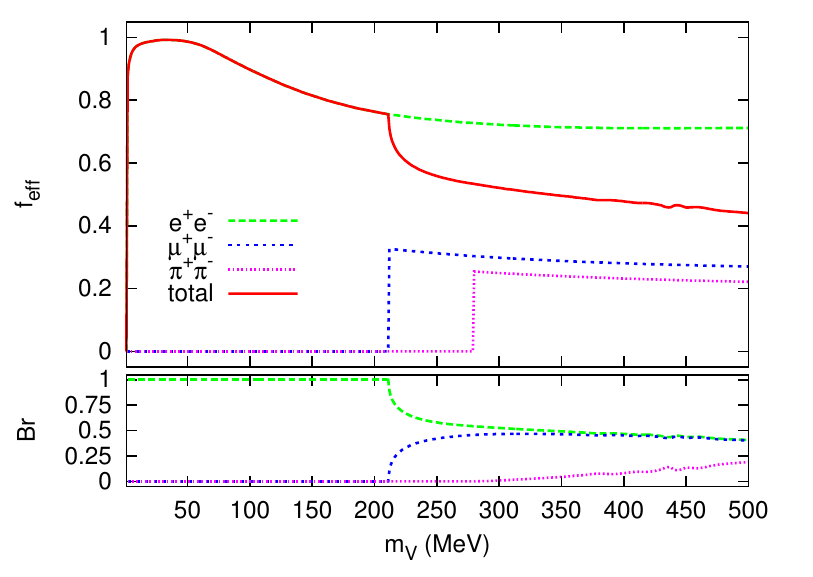}
\caption{Effective deposition efficiency for each decay channel with the sum weighted by the branching ratios for
$\Gamma_V^{-1} = 10^{14}$s.}
\label{fig:feff}
\end{center}
\end{figure}

The energy output $\zeta$ can be related to the VDP parameters as follows,
\be
 \zeta = \frac{f}{3} \frac{\Om_V}{\Om_b} = \frac{f}{3} \frac{E_{\rm p.b.}}{m_p}.
\label{eq:zeta}
\ee
The pre-factor $f$ determines the overall efficiency with which the deposited energy goes into heating and ionization.
The thermalization of an energetic particle depends on the species, initial energy and
redshift~\cite{Padmanabhan:2005es,Slatyer:2009yq}. Ref.~\cite{Slatyer:2012yq} provides transfer functions
$T(z_{inj},z_{dep},E)$ giving the fractional amount of energy deposited at $z_{dep}$ for an energy injection $E$ at
$z_{inj}$ for both $\gamma$ and $e^+e^-$ final states. With this information, we can numerically solve for the
deposition efficiency of the injected energy from decaying particles \cite{Slatyer:2012yq},
\begin{align}
f(z) &= \frac{\left.\frac{d E}{d z}\right|_{\rm dep}(z)}{\left.\frac{d E}{d z}\right|_{\rm inj}(z)} \\
&= \frac{H(z){\displaystyle\sum_\mathrm{species}}\int_{z}^\infty \frac{d \ln(1+z_{\rm in})}{H(z_{\rm
in})} \int
T(z_{\rm in},z,E) E \frac{d \tilde{N}}{d E} d E }{{\displaystyle \sum_\mathrm{species}}\int E \frac{d
\tilde{N}}{d E} d E},
\end{align}
where $\frac{d \tilde{N}}{d E}$ is the normalized energy distribution of the $e^+e^-$ or $\gamma$ in the decaying
particle rest frame. This strategy has been used in Refs.~\cite{Slatyer:2009yq,Cline:2013fm} to analyze dark matter
annihilation and decay to standard model particles for $m_\chi > 1$ GeV. An effective deposition efficiency $f_{\rm
eff}$ is found by averaging $f(z)$ over the range $800 < z < 1000$. We compute $f_{\rm eff}$ for VDP in the mass range
1-500~MeV where the decay channels are $V\to\{e^+e^-,\mu^+\mu^-,\pi^+\pi^-\}$~\cite{Batell:2009yf}. The results for
$f_{\rm eff}(m_V)$, along with each decay channel contributions and their branching ratios, are shown in Fig.~\ref{fig:feff} for
$\Gamma_V^{-1} = 10^{14}$s. The low efficiency of $\mu^\pm$ and $\pi^\pm$ is due to the neutrinos radiating away a
large fraction of the energy. For $e^\pm$ with $E\gtrsim 100$~MeV, the longer cooling time lowers the
efficiency~\cite{Slatyer:2012yq}, which is clearly seen in the $f_{\rm eff}^{e^\pm}$ curve. 

\begin{figure}[htb]
\begin{center}
\includegraphics[width=.5\textwidth]{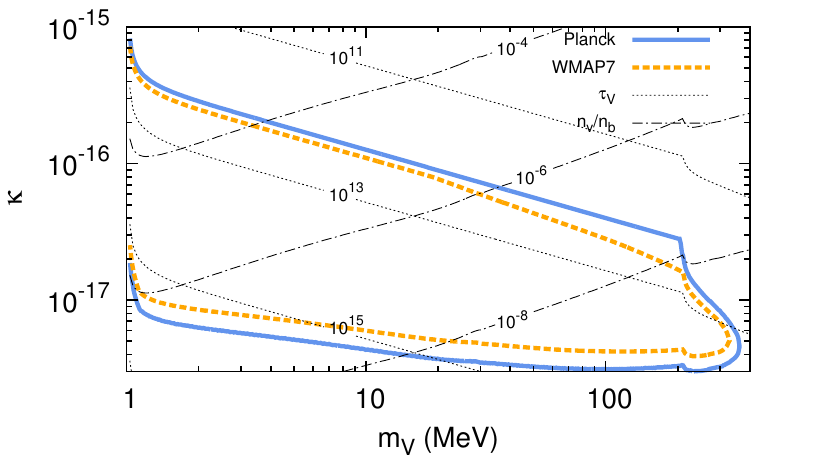}
\caption{The solid contours bound the regions excluded by the CMB constraints on VDP. Contours of the lifetime in seconds and relative number density of dark photons to baryons prior to their decay are also shown.}
\label{fig:cmb}
\end{center}
\end{figure}

Using the result (\ref{Eyield}) with $f_{\rm eff}$ in (\ref{eq:zeta}), we find that our CMB constraints on
$\Gamma-\zeta$ lead to the excluded region of parameter space shown in Fig.~\ref{fig:cmb}. We find this to be a rather remarkable
sensitivity to an effective electromagnetic coupling as small as $\al_{\rm eff} \sim 10^{-37}-10^{-38}$!

\section{Concluding Remarks}

The kinetic mixing portal is one of the few renormalizable interaction
channels between the SM and a neutral hidden sector. As such, it is of
interest to understand the full spectrum of limits on dark photons
coupled through this portal. In this paper, we have determined the
cosmological constraints due to the impact of late decays on BBN and
the CMB; the sensitivity extends to remarkably small effective
electromagnetic couplings.  In this concluding section, we comment on
possible indirect signatures in the present day universe from the
decay of relic dark photons, and other potential extensions.

It is important to emphasize that the constraints derived in this work
rely only on the thermal production of VDP and the minimal cosmological history of the
Universe.  For the mass range of VDP considered here, the
constraints will hold as long as temperatures $T\sim {\cal O}(1) -{\cal O}(100)$ MeV were attained 
at an early epoch. Any additional contributions to the
abundance of VDP, such as production of $V$ through other portals, or nonthermal contributions to $m_V^2 \langle
V_\mu^2\rangle $ due to vacuum misalignement mechanisms, will enhance the VDP abundance, and correspondingly strengthen
the bounds on $\kappa$.

The analysis in this paper assumed that the vector mass was above the electron threshold. For lower
masses, $V$ naturally has a lifetime well in excess of the age of the universe and can play the
role of dark matter \cite{Pospelov:2008jk,Redondo:2008ec}. In this regime its relic abundance is fixed 
instead by Thomson-like scattering, $e+\gamma \rightarrow e +V$. As discussed in \cite{Pospelov:2008jk},
for $m_V\sim 100$~keV, indirect constraints still allow this cosmological abundance with $\ka \sim 10^{-11}$, but photoelectric
absorption in dark matter detectors would leave a detectable ionization signal. The electronic background
data from XENON100 in the 1-100~keV range \cite{Aprile:2011vb} indicated no signal, thus appearing to close this window,
as discussed in more detail in \cite{Arisaka:2012pb}. Very recently, these limits have also been improved by XMASS \cite{Abe:2014zcd}.
Nevertheless, minimal extensions of VDP in this mass range can provide viable models of superweakly-interacting massive particle dark matter. 
One option is to have a dark Higgs $h'$ responsible for breaking $U(1)_V$ and generating the dark photon mass. In the $m_{h'} < m_V$ regime, this will lead to 
extremely long-lived $h'$ particle states since $\Gamma_{h'} \propto \kappa^2$ \cite{Batell:2009yf}. 
In this case, one would require somewhat larger values of $\kappa$ to ensure a more efficient $e^-e^+ \to Vh'$ production channel. Another option is simply a new state 
state $\chi$, which is stable and charged under $V$. The analysis of these very light dark matter models goes beyond the scope of the present paper.

We can also consider a higher mass range, e.g.~TeV-scale dark
photons, whose present-day decays could provide signatures in
antimatter, gamma-ray and neutrino observations~\cite{Ibarra:2013cra}.
With a more massive dark vector, the full kinetic mixing with
hypercharge should be included, $ {\cal L}_{\rm V} =
-\frac{\tilde\kappa}{2} B_{\mu\nu} V^{\mu\nu} = -\frac{\kappa}{2}
F_{\mu\nu} V^{\mu\nu} + \frac{\kappa\tan\theta_w}{2} B_{\mu\nu}
Z^{\mu\nu}$, where $\tilde\kappa= \kappa \cos \theta_w$ to keep the
same normalization as before. Fermions then acquire both vector and
axial vector couplings to $V$, modifying both the production and
decays rates. Assuming $m_V \gg m_Z$, and generalizing (\ref{eq:M2}) by
summing over all degrees of freedom for $\gamma$ and $Z$ mediation,
leads to
\be
\Gamma_V 
  \simeq  10^{-17} \mbox{s}^{-1}\left( \frac{\alpha_{\rm eff}}{10^{-45}} \right) \left(\frac{m_V}{1\mbox{ TeV}}\right) .
\ee
In the MB approximation, freeze-in production is analogous to (\ref{Ysimple}), 
and using $g_\star \simeq 100$ and summing over the channels, we find
\be
Y_{V,f} 
 \simeq 10^{-31} \left( \frac{\Gamma_V}{10^{-17}{\rm s}^{-1}} \right) \left( \frac{\mbox{TeV}}{m_V} \right)^2.
\ee
This is minuscule compared to the cold dark matter energy density
\be
\frac{n_V m_V}{\rho_{\rm CDM}} 
  \simeq 10^{-19} \left( \frac{\Gamma_V}{10^{-17}{\rm s}^{-1}} \right).
\ee
Decaying dark matter of that mass range, with 100\% leptonic
branching, requires a lifetime of $\tau_{\rm DM} =
10^{26}$s~\cite{Ibarra:2013cra} to contribute to the increasing
positron fraction in cosmic rays observed by
PAMELA~\cite{Adriani:2008zr} and AMS-02~\cite{Aguilar:2013qda}. The
VDP scenario thus falls short by many orders of magnitude. Similar
conclusions follow for neutrino experiments, where decaying dark
matter with mass $10 - 10^{15}$ TeV requires a lifetime of ${\cal
  O}(10^{26} -
10^{28})$s~\cite{Esmaili:2012us, Murase:2012xs}. Very long-lived
dark photons are therefore too feebly coupled in this minimal scenario
to contribute to these indirect detection signals.

Finally, we note that the analysis performed in this paper can easily ve extended to other
cases of ``very dark" particles. For example, super-weakly
interacting singlet scalars $S$, coupled to the SM via the
renormalizable Higgs portals $ASH^\dagger H + \lambda S^2 H^\dagger H$ can
be probed via BBN \cite{Pospelov:2010cw} and the CMB. While the
main cosmological constraints will be very similar to the VDP case, the 
details of the production from the Higgs portal are different, and shifted to the
earlier electroweak epoch. The analysis of this minimal scalar model 
is on-going~\cite{us_and_them}.

\subsection*{Acknowledgements}

We would like to thank H.~An, J.~Redondo and D.~Walker for
helpful discussions.  This work was supported in part by NSERC,
Canada, and research at the Perimeter Institute is supported in part
by the Government of Canada through NSERC and by the Province of
Ontario through MEDT. The work of AF is partially supported by the
Province of Qu\'ebec through FRQNT. JP is supported by the New Frontiers 
program of the Austrian Academy of Sciences.

\section*{Appendix A: Degrees of freedom}

Our evaluation of the number of relativistic degrees of freedom needed in the Hubble rate and entropy density follows
the technique used in~\cite{Hindmarsh:2005ix}, updated with more recent theoretical QCD results. 

\begin{figure}[htb]
\begin{center}
\includegraphics[width=.45\textwidth]{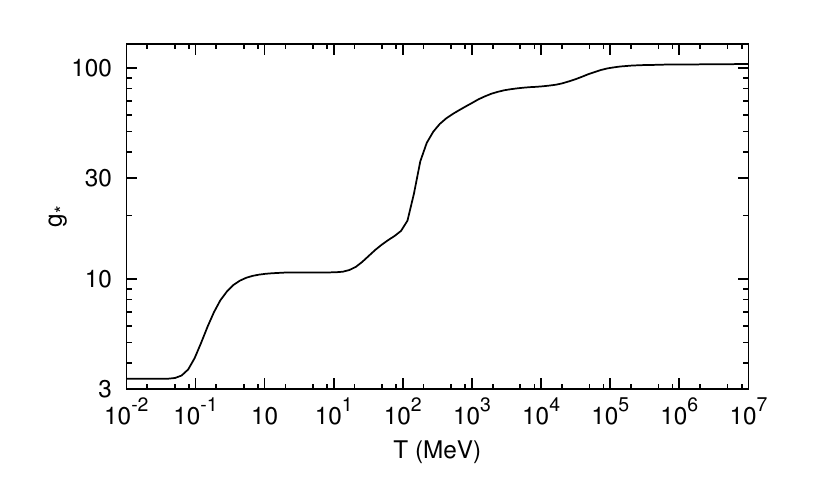}
\caption{Relativistic degrees of freedom as a function of temperature.}
\label{fig:gstar}
\end{center}
\end{figure}

The BMW lattice QCD group~\cite{Borsanyi:2013bia} provides a fitting function for the trace anomaly, from
which we can extract the energy and entropy density. Their function incorporates the hadron resonance gas model
below the pseudo-critical temperature $T_c$ and $n_f = 2+1$ lattice results up to 1000 MeV. At higher temperatures, we
used the $n_f = 3$ three-loop result from hard-thermal-loop perturbation theory~\cite{Haque:2014rua} with
renormalization scale $\Lambda = 2\pi T$. 
The heavier quarks are modelled as an ideal gas, scaled by the ratio of the energy density of $n_f =3$ QCD to the
ideal gas value at the given temperature. This approximation has been used in~\cite{Hindmarsh:2005ix} and is
shown to be in good agreement with preliminary lattice results for $n_f = 2+1+1$~\cite{Borsanyi:2010cj}. The resulting
$g_\star (T)$ is shown in Fig.~\ref{fig:gstar}. 

The QCD phase transition is a cross-over, with a pseudo-critical temperature $T_c$ in the range of 150-170~MeV. For a
given observable, $T_c$ is well-defined as the temperature of the maximal inflection point. In the present work, we used $T_c
= 157 $~MeV, the pseudo-critical temperature of the energy density~\cite{Borsanyi:2010bp}.

\section*{Appendix B: Resonant Production}

Here we demonstrate that the thermal effects, and the associated resonant production, create a parametrically suppressed 
contribution to $Y_{V,f}$, although numerically it may constitute as much as 30\%. 

\subsection*{Relativistic Case}

We begin the analysis by choosing the 
simplest case of electron-positron coalescence and use MB statistics. Since thermal effects are going to be important 
at higher temperatures than $m_V$,  $m_e$ is negligible and can be set to 0 from the start.
Furthermore, we break up  the matrix element into the longitudinal and transverse pieces according to the polarization of the 
$V$ boson produced with four-momentum $(\omega, \vec{q})$ to derive the right-hand side of the Boltzmann equation (\ref{Beq}).
After direct calculation we obtain
\begin{eqnarray}
{\rm R.H.S.} = \frac{3}{2\pi^2} m_V\Gamma_{V\to e\bar{e}}\int_{m_V}^\infty d\omega\sqrt{\omega^2-m_V^2} e^{-\omega/T}
\nonumber\\
\times \left\{ \fr13 \frac{m_V^4}{|m_V^2- \Pi_L |^2}+\fr23 \frac{m_V^4}{|m_V^2- \Pi_T|^2} \right\}.~~~~~
\label{thermal}
\end{eqnarray}
The polarization tensors $\Pi_{T(L)}$ are complex functions of $\omega, |\vec{q}|$ and $T$, and originate from the virtual photon propagators. In the limit of vanishing plasma density, $\Pi_{T(L)}\to0$, the expression inside $\{...\}$ tends to 1, and the 
R.H.S. becomes identical to that of (\ref{simplified}), as it should be. The expressions for $\Pi_{T(L)}$ can be found in the thermal field theory literature, and we use the results of \cite{Braaten:1993jw}, with the more symmetric 
definition of the longitudinal polarization tensor \cite{An:2013yfc}, 
$\Pi_L^{\rm this\,work} = \frac{m_V^2}{\omega^2-m_V^2}\Pi_L^\text{Ref.\,\cite{Braaten:1993jw}}$.

\begin{figure}
\begin{center}
\includegraphics[width=.45\textwidth]{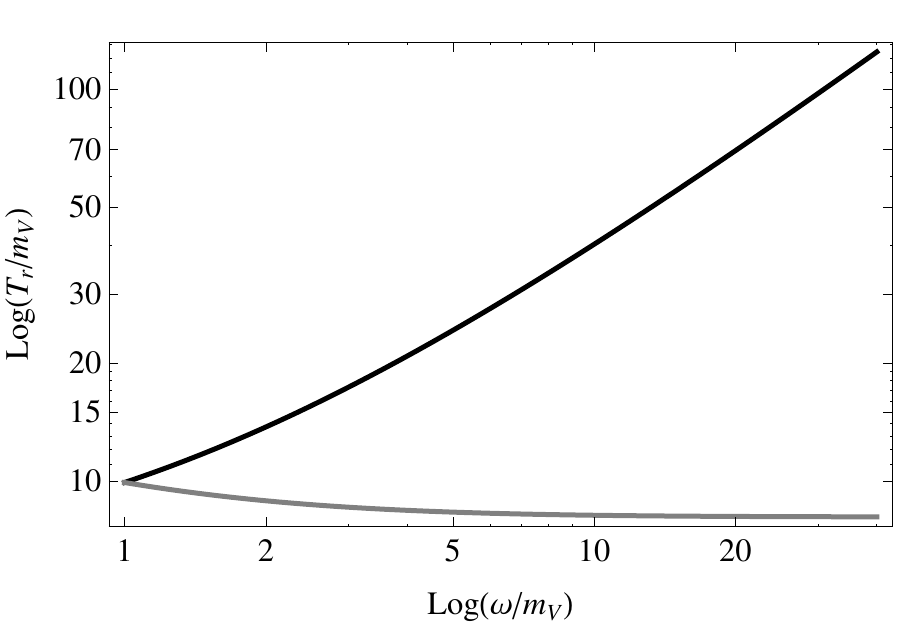}
\caption{The dependence of the resonant temperatures $T_{r,L}$ (black) and 
 $T_{r,T}$ (gray) on frequency $\omega$, 
all in units of $m_V$. The transverse resonance frequency asymptotes to $T_{\rm min} = m_V(3/(2\pi\alpha))^{1/2}$. }
\label{FigRes}
\end{center}
\end{figure}

For a one-component ultra-relativistic plasma (again neglecting muon and pion contributions etc.),
the expressions for the real parts of the polarization tensors are given by \cite{Braaten:1993jw},
\begin{eqnarray}
{\rm Re}\Pi_T(\omega) &=& \omega_p^2\frac{3\omega^2}{2\vec{q}^2}\left( 1-\frac{m_V^2}{\omega^2}\frac{\omega}{2|\vec{q}|}
\log\frac{\omega +|\vec{q}|}{\omega -|\vec{q}|}\right), \nonumber
\\
{\rm Re}\Pi_L(\omega) &=& 3\omega_p^2\frac{m_V^2}{\vec{q}^2}\left( \frac{\omega}{2|\vec{q}|}
\log\frac{\omega +|\vec{q}|}{\omega -|\vec{q}|} - 1 \right), \label{eq:RePi}
\end{eqnarray}
where all the factors of $|\vec{q}|$ can be replaced with $\sqrt{\omega^2-m_V^2}$. 
The plasma frequency of the electron-positron fluid  is given by 
\be
\omega_p^2 = \frac{4\pi\alpha}{9}T^2.
\ee
The imaginary parts of the polarization tensors are related to 
the massive photon decay rate in vacuum, $\Gamma_0 = \alpha m_V/3$, 
\be
{\rm Im \Pi}_{T(L)} = -\Gamma_0 m_V\left(1-\exp(-\omega/T)\right).
\ee
(The VDP decay rate in vacuum is $\kappa^2\Gamma_0$ in this approximation.) Armed with these expressions, we can derive the 
conditions for a resonance, that is the point in $\{T,\omega\}$ where the denominator of (\ref{thermal}) is minimized,
\begin{eqnarray}
{\rm Re}\Pi_{T(L)}(\omega, T_{r,T(L)}) = m_V^2.
\end{eqnarray} 
The dependence of $T_{r,T(L)}(\omega)$ is plotted in Fig. \ref{FigRes}. The most important point is that all resonance 
frequencies are parametrically larger than $m_V$, and there is a minimum 
frequency at which the resonance can happen,
\be
\label{Tmin}
T_{\rm min}  = m_V \left[ \frac{3}{2\pi\alpha}\right]^{1/2} \simeq 8.1 \,m_V.
\ee
Thus all resonances occur at temperatures that are parametrically larger (by a factor of $\alpha^{-1/2}$) than $m_V$, where the 
Hubble expansion rate is significantly greater than at $T< m_V$. We proceed by calculating the resonant contributions 
by using the narrow width approximation, that is we approximate the ratios inside the $\{...\}$ of Eq.~(\ref{thermal})
by delta functions, 
\be
\frac{m_V^4}{|m_V^2- \Pi|^2} \simeq \frac{3 \pi}{2 \alpha}\, \frac{T_r(\omega) \delta[T-T_r(\omega)]}{e^{\omega/T}-1}.
\ee
This expression holds for both the $T$ and $L$ resonances. 

The resonant contribution to the VDP abundance comes from evaluating two integrals, over $T$ and $\omega$. 
If the integral over the temperature is performed first, one finds
\begin{eqnarray}
\Delta Y_{f,r} &=& Y_T +Y_L,~~~~~~~~~~~~~~~~\\
Y_{T(L)} &=& \frac{3g_{T(L)}}{4\pi\alpha} \frac{m_V^3\Gamma_{V\to e\bar{e}} }{(Hs)_{T=m_V}}\int_{m_V}^\infty
\frac{m_V^3\sqrt{\omega^2-1} d\omega}{(T_{r,T(L)}(\omega))^5[e^{\omega/T}-1]},
\nonumber
\end{eqnarray}
where $g_{T(L)}=2(1)$ are the multiplicity factors. Performing the remaining integral 
we arrive at the following result,
\be
\label{MBres}
\Delta Y_{f,r}\simeq \Delta Y_T \simeq 0.17 \times
\frac{m_V^3\Gamma_{V\to e\bar{e}} }{(Hs)_{T=m_V}}.  \ee The
longitudinal resonance turns out to be negligible on account of the
large value of $T_{r,L}$ when $\omega\sim T_{r,L}$.  (This is in
contrast with the stellar production of very light dark photons, where
the $L$-resonance dominates \cite{An:2013yfc}.)  We now see that
although the resonant contribution (\ref{MBres}) is parametrically
suppressed, by $O(\alpha^{1/2})$, relative to the continuum
contribution (\ref{Ysimple}), it can reach $20\%$ of the
total. Repeating the same calculations with FD statistics changes the
coefficient only slightly, $0.17_{\rm MB} \to 0.15_{\rm FD}$.

\subsection*{Nonrelativistic Corrections}

The analytical treatment of  resonant production above  is only valid for massless particles in the loop. In our numerical
calculations, we include the $\Pi_{T,L}$ effects for all leptons of mass $m<10$~GeV and charged pions for $T<T_c$.
Ref.~\cite{Braaten:1993jw} provides analytical approximations for Re$\Pi_{T,L}$, which interpolate smoothly between
the `classical' (nonrelativistic) and relativistic limits, 
\begin{align}
\rm{Re}\Pi_T &= \omega_p^2 \frac{3}{2v_\star^2}\left(\frac{\omega^2}{k^2} - \frac{\omega^2-v_\star^2
k^2}{k^2}\frac{\omega}{2 v_\star k}\log\frac{\omega + v_\star k}{\omega - v_\star k}\right), \label{eq:RePiT}\\
\rm Re \Pi_L &= \omega_p^2 \frac{3m_v^2}{v_\star^2 k^2}\left( \frac{\omega}{2v_\star k} \log \frac{\omega + v_\star
k}{\omega - v_\star k} - 1\right) \label{eq:RePiL},
\end{align}
where 
\begin{align}
\omega_p^2 &= \frac{8\alpha}{\pi} \int_0^\infty dp \frac{p^2}{E}\left(1-\frac{1}{3}\frac{p^2}{E^2}\right)n_F
(E),\\
 w_1^2 &=  \frac{8\alpha}{\pi} \int_0^\infty dp
\frac{p^2}{E}\left(\frac{5}{3}\frac{p^2}{E^2}-\frac{p^4}{E^4}\right)n_F (E), \\
v_\star &= \frac{w_1}{w_p}.
\end{align}
The parameter $v_\star$ can be interpreted as the typical velocity of the fermion at that given energy. We recover the
relativistic limit~(\ref{eq:RePi}) with $v_\star \to 1$ and the `classical' (nonrelativistic) limit with
$v_\star \to \sqrt{5T/m_f}$.

In general, the imaginary part of the polarization tensor is given by~\cite{Weldon:1983jn},
\begin{align}
{\rm Im} \Pi &= -\omega \Gamma^{\rm{Prod}}\left(e^{\frac{\omega}{T}}-1\right),\\
\Gamma^{\rm Prod} &= \frac{1}{2\omega} \int \frac{d^3p}{2E_p} \frac{d^3q}{2E_1} \frac{(2\pi)^4}{(2\pi)^6} \delta^4(k-p-q)  \\ &\qquad \qquad\qquad  \times \left|\mathcal{M}_{1,2\to V}\right|^2 n_1 n_2.
\end{align}
Here $\Gamma^{\rm Prod}$ represents the production rate, with $\mathcal{M}_{1,2\to V}$ the matrix element for the particles coalescing into $V$ and $n_1$ and $n_2$ their respective statistical distributions. Separating the T and L parts of matrix element,
\begin{align}
\left|\mathcal{M}_{l\bar{l}\to V}^{T,L} \right|^2 
&= 16 \pi \alpha (F^T + F^L),  \\
F^T &= -2p^2\sin^2\theta+m_V^2, \\
F^L &= -\frac{2}{m_V^2}\left(kE_p-\omega p \cos\theta\right)^2 + \frac{m_V^2}{2},
\end{align}
we find
\begin{align}
\Gamma_{T(L)}^{\rm Prod} &= \frac{\alpha}{\omega k} \int_{\frac{\omega}{2}-\frac{k}{2}\sqrt{1-4\frac{m_f^2}{m_V^2}}}^{\frac{\omega}{2}+\frac{k}{2}\sqrt{1-4\frac{m_f^2}{m_V^2}}} dE_p\; \times \\ &\qquad \qquad \times  F^{T(L)}(\omega, p, \theta)\; n(E_p)n(\omega - E_p),
\label{eq:ImPi-prodrate}
\end{align}
where $k$ relates to the dark vector, $p/q$ to the fermions in the loop and $\cos\theta = \frac{\omega
E_p}{kp}-\frac{m_V^2}{2pk}$.

\section*{Appendix C: Hadronic Production}

To model hadronic freeze-in production, we treat the coalescence of charged pions into dark photons as a scalar QED process. The spin-summed matrix element is
\begin{equation}
\sum |\mathcal{M}_{s\bar{s}}|^2 = 4\pi \alpha_{\rm eff}^{\pi\pi} m_V^2 \left(1-4\frac{m_s^2}{m_V^2}\right),
\end{equation}
with the massless limit being a factor of 4 smaller than the fermionic case~(\ref{eq:M2}). We include the
$\rho$-resonance in the charged-pion interaction via an effective scalar electromagnetic coupling which becomes $m_V$
dependent, $\alpha_{\rm eff}^{\pi\pi}(m_V) = \kappa^2 
\alpha^{\pi\pi} (\sqrt{s} = m_V)$. The coupling function $\alpha^{\pi\pi} (\sqrt{s})$ is extracted numerically from the $e^+e^- \to
\gamma^* \to \pi^+\pi^-(\gamma)$ cross section measured by BaBar collaboration~\cite{Lees:2012cj}, and similarly for the charged kaons~\cite{Lees:2013gzt}.

In accounting for thermal effects, the imaginary part of the polarization tensor can be found in the same manner as in Appendix B, by separating the matrix element into the different propagation modes for scalars,
\begin{align}
F^T_s &= p^2\sin^2\theta , \\
F^L_s &= \frac{1}{m_V^2}\left(kE_p-\omega p \cos\theta\right)^2,
\end{align}
and~(\ref{eq:ImPi-prodrate}) can be used with Bose-Einstein statistics.

The real part of the polarization tensor needs to be derived from first principles in  finite-temperature field theory as the $\omega/k$ scaling of~(\ref{eq:RePiT})~(\ref{eq:RePiL}) does not generally hold. However, it is known~\cite{Kapusta:2006pm} that the high temperature limit is the same as~(\ref{eq:RePi}), since the statistics integrals $\int_0^\infty dp\; p\, n_B(p) = 2\int_0^\infty dp\; p\, n_F(p)$ compensate for the missing spin degrees of freedom~\cite{Rychkov:2007uq}. On account of the high resonant temperature~(\ref{Tmin}), we find that we can maintain good numerical accuracy with the simple rescaling,
\be
{\rm Re\Pi^s_{T(L)}} = \frac{\rm{Re}\Pi_{T(L)}}{2}\frac{\int dp\; \frac{p^2}{E}n_B(E)}{\int dp\; \frac{p^2}{E}n_F(E)}.
\ee

\section*{Appendix D: BBN Analysis}

Here we provide some additional details regarding the treatment of
BBN; the analysis of meson injection draws in large parts from our
previous paper~\cite{Pospelov:2010cw} to which we refer the reader for
an exhaustive discussion. The Boltzmann code that we use is based on
Ref.~\cite{Kawano:1992ua}, but incorporates some
significant improvements and updates. These are likewise detailed
in~\cite{Pospelov:2010cw}. Our SBBN yields are in excellent agreement
with those presented in~\cite{Cyburt:2008kw} at the WMAP value of
$\eta_{b}= 6.2\times 10^{-10}$ and with a neutron lifetime of
$\tau_n=885.7\,$s.

Below the di-pion threshold, $m_V \leq 2 m_{\pi^{\pm}} = 279\,\MeV$,
only electromagnetic energy injection from $V$-decays is relevant. As
discussed in Sec.~\ref{sec:impact-bbn}, the formation of a photon
cascade $f_{\gamma}(E_{\gamma})$ gives way to photodissociation of
nuclei.
The rate of destruction of a species $N$ with number density $n_N$ is
then given by
\begin{align}
  \Gamma_{\mathrm{ph}} (T) = 2 n_N \int_{E_{\mathrm{thr}}}^{E_{\mathrm{max}}}
  dE_{\gamma} \,  f_{\gamma}(E_\gamma) \sigma_{\gamma + N
    \to X}(E_\gamma),
\end{align}
where $\sigma_{\gamma + N \to X}(E_\gamma)$ is the photo-dissociation
cross section for $\gamma + N \to X$ with threshold
$E_{\mathrm{thr}}$. The factor of two accounts for the
back-to-back $e^{\pm}$ pair forming two independent cascades, each with
a maximum energy of $E_{\mathrm{max}} = \max \left\{
  E_{\mathrm{pair}}, E_{\mathrm{inj}}/2 \right\}$. We take into
account all relevant light element reactions listed
in~\cite{Cyburt:2002uv} and we also include secondary processes
which may result in production of~\lisx. The Boltzmann equations
describing the temperature evolution of the light elements in the
presence of energy injection are straightforward to obtain.

With regard to the injection of mesons and nucleons, we restrict
ourselves to reactions at threshold, assuming that charged pions and
kaons are thermalized before reacting. Likewise we assume that
neutrons will be slowed down by their magnetic moment interaction with
electrons, positrons and photons and neglect neutral kaons altogether because of their
inability to stop and the associated uncertainty in reaction cross section.

We expect such an approximation to result in more conservative
constraints. Incomplete thermalization for charged mesons only
happens on the whole for temperatures $T< 40\,\keV$, for which the plasma stopping
power diminishes. Away from threshold, pion-nucleon reactions can
proceed resonantly, \textit{e.g.}~$ \pi^- p \to \Delta^0 \to \pi^0 n $,
with an efficiency up to $\sim 20-30$ times the value for stopped
pions. Likewise, the total inelastic $\pi^-$-\hef\ cross section
becomes significantly larger for pion kinetic energies of $\sim
150\,\MeV$. Such enhancements as well as non-thermal neutrons with
spallating power lead to stronger departures from the standard case
and are therefore more strictly constrained.
There is, however, the beneficiary effect of reducing the cosmological
lithium abundance towards observationally favored values through the
production of ``extra neutrons''. As pointed out
in~\cite{Pospelov:2010cw}, this process can also be boosted by the
above resonances.  However, this solution of the lithium problem is
challenged by the simultaneous tightening of the D/H constraint,
especially in light of the new D/H determinations discussed in the BBN
section.  For the interested reader, we point out that a detailed
quantitative discussion of incomplete stopping can be found in our
preceding work~\cite{Pospelov:2010cw}.

Finally, baryon/anti-baryon pairs can be produced directly in the
decay of the vector for $m_V\gtrsim 2\,\GeV$. Upon injection,
resonances and hyperons decay to (anti)protons and
(anti)neutrons---possibly accompanied pions and kaons---before
interacting with the ambient medium. The fate of the final state
nucleons is then as follows: $\bar n$ and $\bar p$ will preferentially
annihilate on protons which are the most abundant target in the
Universe with an annihilation cross section $\langle \sigma_{\rm
  ann}v\rangle \sim m_{\pi^{\pm}}^{-2}$.  Depending on the $n/p$
ratio, they also annihilate with neutrons with a similar cross
section. The annihilation on protons is faster than the Hubble rate at
all relevant temperatures and---if annihilating on protons---the
injection of $n\bar n $ results in one net $p\to n$ conversion with
associated energy injection of $m_p+m_n$. Likewise, if annihilating on
neutrons, $p\bar p $ injection results in one net $n\to p$
conversion. Assuming equal cross sections, the relative efficiencies
for those processes are $p/(n+p) $ and $n/(n+p) $ respectively and
we treat this sequence of events as being instantaneous.

\begin{figure}[tb]
\begin{center}
\includegraphics[width=0.95\columnwidth]{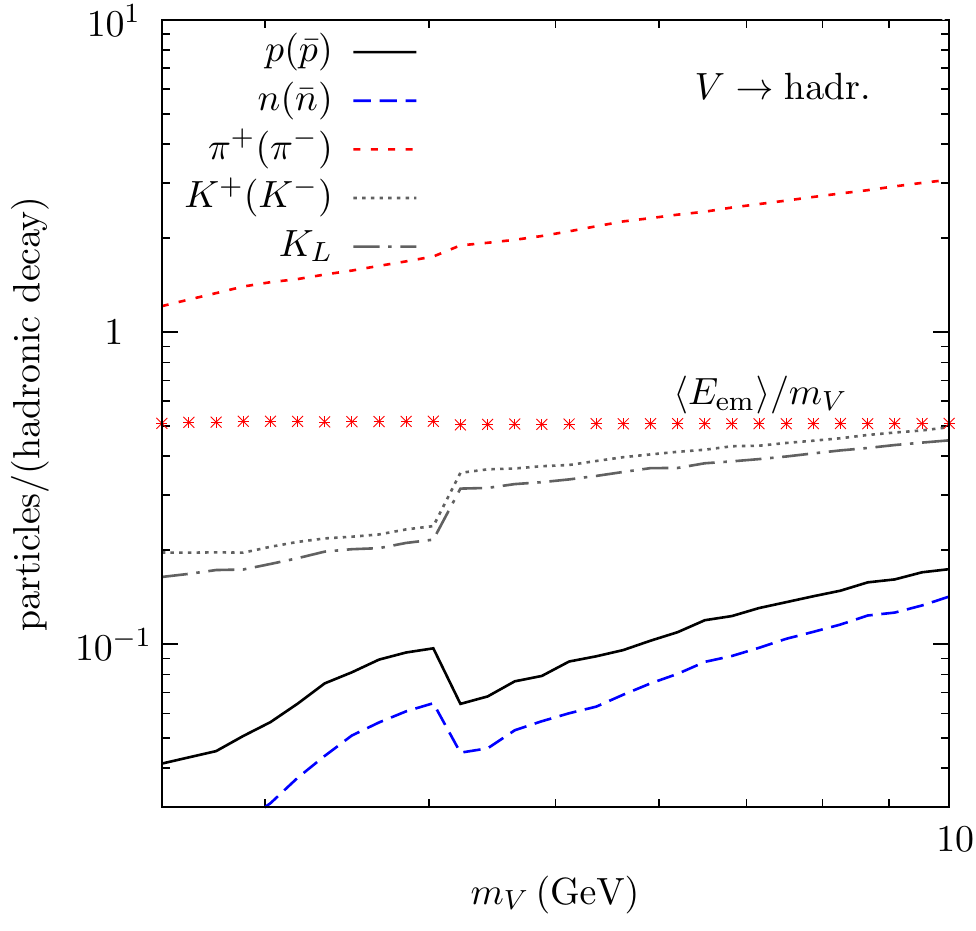}
\caption{The average number of particles per $V$ decay with $m_V > 2.5\,\GeV$, from a Pythia simulation. 
 Also shown is the average
  electromagnetic energy injected after all particles have decayed to
  electrons and photons ($e^+$ are assumed to have annihilated on
  $e^-$.) When including leptonic channels, to a good approximation
  $1/3$ of the energy is carried away in the form of neutrinos. Resonances
  like $J/\psi$ are not captured by the resolution of the simulation and
  we neglect such isolated points in the parameter space. }
\label{fig:pythia}
\end{center}
\end{figure}

Neutron injection during BBN in the decay $V\to n\bar n$ and close to
the threshold $m_V\gtrsim 2m_n$ can be studied by utilizing the (only)
measurement of electron-positron annihilation to the
neutron-antineutron final state, $e^+ e^- \to n \bar n$~\cite{Antonelli:1998fv}.
At threshold, $ \sigma_{e^+ e^- \to n \bar n } \sim 1\,\rm{nb} $ is
reported. With a total hadronic cross section $ \sigma_{e^+ e^- \to
  \rm had } \sim 50\,\rm{nb} $ this points to a branching fraction
$\sim 2\%$.
In our actual analysis we use a more conservative value that arises
from a joint extraction of the neutron Sachs electric (magnetic) form
factor $|G_{E(M)}^n(q^2)| $ in the time-like and space-like regions;
for us, the momentum transfer is time-like with $q^2 = m_V^2$ and
\begin{align}
  \sigma_{e^+ e^- \to n \bar n } & = \frac{4\pi \alpha^2}{3 q^2}
  \sqrt{1 - \frac{4 m_n^2}{q^2}} \nonumber \\ & \times \left[
    |G_M^n(q^2)|^2 + \frac{2 m_n^2}{q^2} |G_E^n(q^2)|^2 \right] .
\end{align}
At threshold we use the solid black line of Fig.~11
of~\cite{Lomon:2012pn} and the $V$-width is then given by
\begin{align}
  \Gamma_{V\to n \bar n} = \kappa^2\frac{m_V^3}{4\pi \alpha} \sigma_{e^+ e^-
    \to n \bar n }(q^2 = m_V^2) .
\end{align}

\begin{figure}[tb]
\includegraphics[width=0.95\columnwidth]{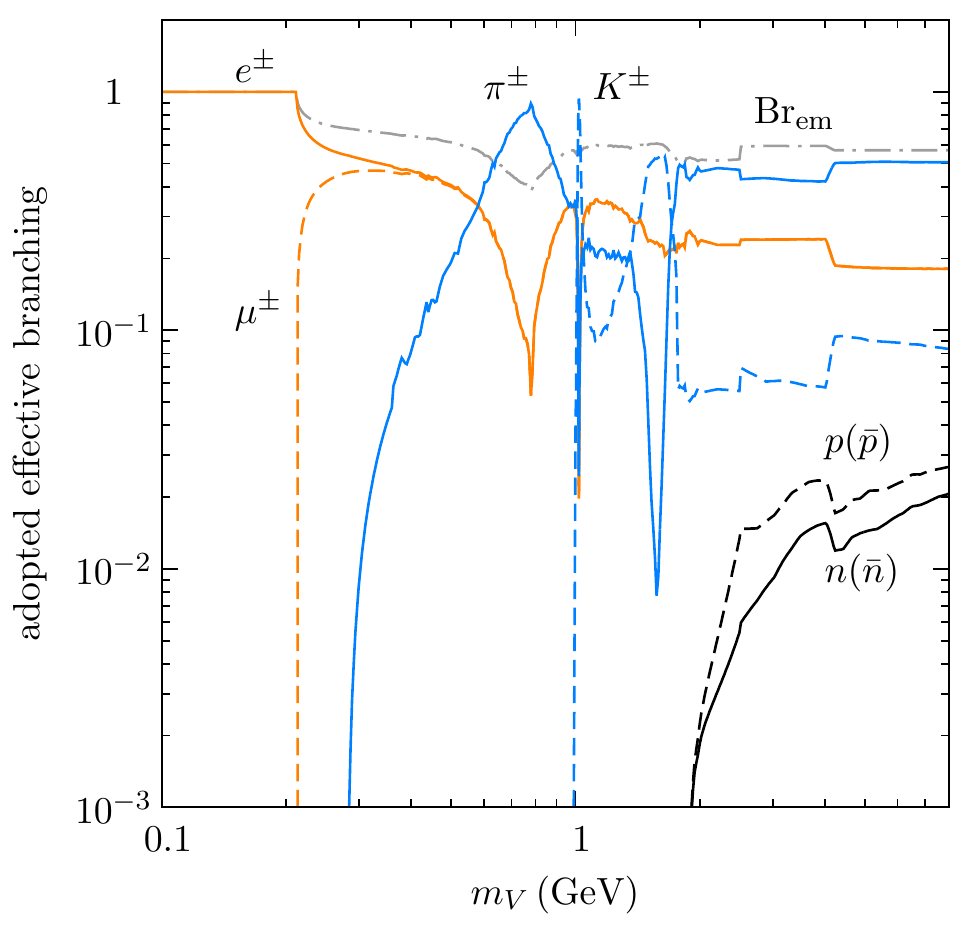}
\caption{The adopted  effective branching ratios into the various final states
  that are relevant for BBN considerations. As multi-pion and kaon
  states become relevant in the kinematically allowed region, we
  stitch together BaBar measurements of the $e^{\pm }\to\pi^{\pm}$ and
  $e^{\pm }\to K^{\pm}$ cross sections up to $m_V = 1.8\,\GeV$ with our
  Pythia simulation for $m_V \geq 2.5\,\GeV$. In this plot, any
  branching to $K_L$ was neglected. Also shown is the fraction of
  vector mass that is converted into EM-energy, denoted Br$_{\rm
    em}$. }
\label{fig:Vdec}
\end{figure}

Away from threshold, we simulate the complex decays of $V$ with
Pythia. In particular, multi-pion(kaon) production and
decays to hyperons and baryonic resonances become relevant.
The yield of phenomenologically relevant final states, $\pi^{\pm},
K^{\pm}$, $K_L$, and nucleons is shown in Fig.~\ref{fig:pythia} for
$m_V\geq 2.5\,\GeV$. Narrow resonances like $J/\psi$ are not captured
by the resolution of the simulation. The dots in Fig~\ref{fig:pythia}
show the average electromagnetic energy injected after all particles
have decayed to electrons and photons; $e^+$ are assumed to have
annihilated on $e^-$. One can see that to a significant fraction of
the energy is carried away by neutrinos.

At lower energies, even though the Pythia simulation is not available,
the topology of the decay events becomes simpler, and is eventually 
dominated by two body decays. Above the di-pion (di-kaon)
threshold, we therefore use BaBar precision measurements of the
$e^{\pm }\to\pi^{\pm}$ and $e^{\pm }\to K^{\pm}$ cross section until
an energy $\sqrt{s} = m_V = 1.8\,\GeV$. Above that energy we stitch the data
together with our simulation above, expecting to capture the overall
importance of various final states qualitatively correctly. The
resulting effective effective branching ratios that are relevant for BBN
considerations are shown in Fig.~\ref{fig:Vdec}. For simplicity, and
as alluded to above, we neglect the $K_L$ contribution.  Also shown is
the fraction of vector mass that is converted into EM-energy in the
hadronic decay, denoted by Br$_{\rm em}$.

\end{document}